\documentclass{amsart}

\usepackage[english]{babel}
\usepackage[utf8x]{inputenc}
\usepackage[T1]{fontenc}

\usepackage[a4paper,top=3cm,bottom=2cm,left=3cm,right=3cm,marginparwidth=1.75cm]{geometry}

\usepackage{amsmath}
\usepackage{amssymb,latexsym,amsmath}
\usepackage{graphicx}
\usepackage[colorinlistoftodos]{todonotes}
\usepackage[colorlinks=true, allcolors=blue]{hyperref}

\newtheorem{theorem}{Theorem}

\begin{document}

\title{On  Some Nonstandard  Heat Transfer Problems}
\author{Ashot Djrbashian,  Armen Arakelyan }
\begin{abstract}
In this article we are making an attempt to connect the theory of functions integrable in the unit disk to the problems of steady-state heat transfer in cases when the heat source is inside the disk.     

\end{abstract}

\maketitle

\section{Introduction}

Let $f(z)$ be analytic in the unit disk $D=\{z: |z|<1\}$ and for $0<p<\infty, -1<\alpha<\infty$
\begin{equation}
||f||^p_\alpha=\int_D|f(z)|^p(1-|z|)^\alpha dm_2(z)<\infty,
\end{equation}
where $dm_2(z)=dxdy$ is the plane Lebesgue measure. Then we say that $f$ belongs to the class $A^p_\alpha=A^p_\alpha(D)$. The theory of $A^p_\alpha$ spaces was established in 1940's by M.M.Djrbashian (see [3] or Djrbashian, Shamoian [2]) and by now is one of the most advanced branches of the theory of Banach spaces of analytic functions. The spaces $A^p_\alpha$ are natural generalizations of the theory of Hardy spaces $H^p, 0<p\leq\infty$ which itself was establish in 1910's by the efforts of multiple mathematicians. In the groundbreaking paper [3] M.M.Djrbashian proved the following integral representation formula:

\begin{theorem}
Let $f\in A^p_\alpha, \; 1\leq p<\infty, \; -1<\alpha<\infty$. Then for $z=re^{i\theta}, \; w=\rho e^{i\phi}$ we have the following integral representation: 
\begin{equation}
f(z)=\frac{1+\alpha}{\pi}\int_0^1\int_{-\pi}^\pi(1-\rho^2)^\alpha\frac{f(\rho e^{i\phi})}{(1-z\rho e^{-i\phi})^{2+\alpha}}\rho d\rho d\phi
\end{equation}
\end{theorem}

As we have mentioned above the mathematical theory of these classes is well known for long time and may be found in the monographs Djrbashian, Shamoian [2], Hadenmalm, Korenblum, Zhu  [6], or Duren, Schuster  [5] (in the special case $\alpha=0$ only). However, unlike the theory of Hardy spaces, as far as we know there are no applications of this theory to concrete physical or engineering problems. In this paper we are doing an attempt to connect the formula (2) in its most basic case to some heat transfer problems. For that purpose we are considering only the simplest case $p=2, \alpha=0$ to make it accessible to widest possible audience of applied mathematicians and engineers with the hope that this approach will promote some additional research. Our idea is that the formula (2) is "too good" to not have useful applications. In some sense we can say, paraphrasing the title of a famous play, that this is  "A formula in search of the problem".

\section{Some integral kernels}
\label{sec:headings}

In this article we will be dealing with harmonic functions  defined in the unit disk $D$. By definition, a function $f(x,y)$ is harmonic if
$$\Delta f(x,y)=\frac{\partial^2 f}{\partial x^2}+\frac{\partial^2 f}{\partial y^2}=0.$$
In polar coordinates $x=r\cos\theta, y=r\sin\theta$ the above equation has the form
$$\frac{\partial^2 f}{\partial r^2}+\frac{1}{r}\frac{\partial f}{\partial r}+\frac{1}{r^2}\frac{\partial^2 f}{\partial\theta^2}=0.$$

Poisson kernel for the unit disk $D$ is the function (in polar coordinates)
\begin{equation}
 P_r(\theta)=P(r,\theta)=\frac{1-r^2}{1-2r\cos\theta+r^2}.
 \end{equation}
It is a long established fact that this kernel solves the so-called classical Dirichlet problem for Laplace's equation in the disk: Find a harmonic function $u$ inside the disk  that on the boundary $T=\partial D$ has the prescribed continuous values $f$. The solution is just the convolution of that function $f$ and the Poisson kernel:
\begin{equation}
u(r,\theta)=\frac{1}{2\pi}\int_{-\pi}^{ \pi}f(\varphi)P_r(\theta-\varphi)d\varphi.
\end{equation}
This solution in effect says that in order to know the values of any harmonic function inside the disk it is sufficient to know its values on the circle, the boundary of the disk.

In fact this formula extends to much wider class of boundary functions $f(\theta)$ than continuous functions. For example, if $[a,b]\subset [-\pi,\pi]$ and $f$ is the characteristic function of that interval then formula (2) still produces a harmonic function $u$ that has boundary values equal to $f$ (in some more general sense). For our purposes we will assume  that $f\in L^2(T)$, the Lebesgue space of square integrable  functions on the unit circle. The resulting function $u$ will belong to harmonic Hardy class $h^2$ satisfying the condition

$$\sup_{0\leq r<1}\int_{-\pi}^\pi |f(re^{i\theta})|^2 f\theta<\infty.$$

The Poisson kernel (3) is one of many examples of  "reproducing kernels". Among the most important properties of the Poisson kernel is the fact that  $P_r(\theta)\geq0$ and for any $r, 0\leq r<1, \frac{1}{2\pi}\int_{-\pi}^{\pi}P_r(\theta)d\theta=1.$ Below you can see the graph of the kernel $P$ for the values  $r=0.5, 0.75,$ and  $0.85$.

\centerline{\includegraphics[width=4in]{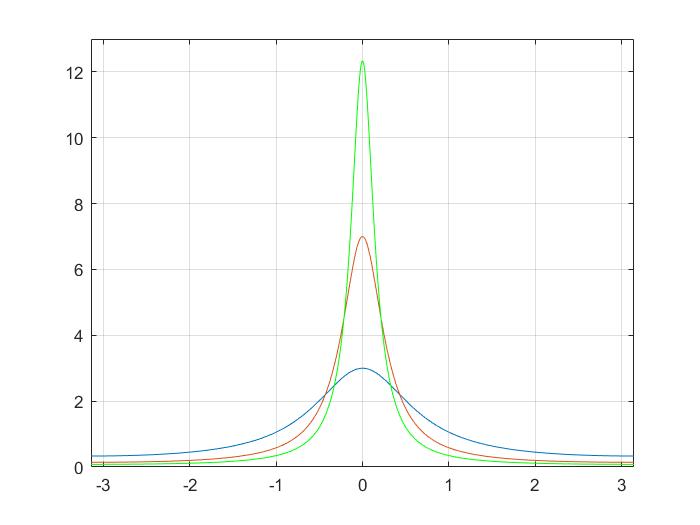}}

$$Figure \; 1$$

The term reproducing kernel is used to indicate the property of reproducing a given harmonic function from its boundary values. We will return to this discussion a little later when we introduce the next integral kernel for harmonic functions.

All the facts about Poisson kernels and Poisson integrals are so widely known that could be found in virtually any Complex Analysis or PDE textbook. We refer interested reader to the book  by Duren [4], for example. 

Next let us consider a different type of integral kernel and integral representation for harmonic functions in the disc. Here the functions are taken from much wider class satisfying condition

\begin{equation}
    \int_{D}|u(x,y|^2dxdy<\infty.
\end{equation}

Here integration is over the unit disk with respect of regular plane Lebesgue measure and we will denote the class of such functions by $hA^2=hA^2(D)$. It is widely known that unlike harmonic functions from the Hardy class $h^2$ functions from $hA^2$ generally speaking do not have boundary values. That means (again, generally speaking) that no representation with Poisson kernel is possible for functions of the class $hA^2$. For these and many other facts about these classes see, e.g. Djrbashian , Shamoian [2] or Axler, Bourdon, Ramey  [1]. The following theorem was never formulated as a separate result anywhere but can be easily deduced from much more general theorems 1.2 or 7.1 from [2].
\begin{theorem} 
If $u\in hA^2$ then it has the integral representation
\begin{equation} 
u(r,\theta)=-u(0)+\frac{2}{\pi}\int_0^1\int_0^{2\pi}u(\rho,\phi)\frac{1-2r\rho\cos(\theta-\phi)+r^2\rho^2\cos2(\theta-\phi)}{(1-2r\rho\cos(\theta-\phi)+r^2\rho^2)^2}d\phi \rho d\rho
\end{equation}
\end{theorem}

Even though the kernel function 

\begin{equation}
  Q(r,\theta)=  \frac{1-2r\cos\theta+r^2\cos2\theta}{(1-2r\cos\theta+r^2)^2}
\end{equation}
does not formally reproduce the function $u(r,\theta)$ (we still need to subtract the value of $u$ at the origin) we will call it  reproducing kernel for the space $hA^2$ and in the future disregard the first term in formula (6) unless it is absolutely necessary. 

By taking $u(r,\theta)=1, 0\leq r<1, 0\leq\theta<2\pi$ we easily see that 

$$\frac{1}{\pi}\int_0^1\int_0^{2\pi}Q(r\rho,\theta-\phi)\rho d\theta d\rho =1$$
for all values of $r, \theta$.

It is also important to notice that in order to "recover" any given harmonic function from that class we need the values of that function on every point in the disk. That is the crucial difference between Poisson representation (4) and representation with the kernel $Q$.

In what follows we will establish some important properties of the kernel function $Q(r,\theta)$. First of all notice that it is not non-negative and that will be demonstrated below with the graph of the function for different values of $r$.  However many important properties are still valid here, starting of course with its reproducing kernel property, expressed by the formula (6).

We start with the graphs of the kernel $Q$ for the values $r=0.5, 0.75$

\centerline{\includegraphics[width=4in]{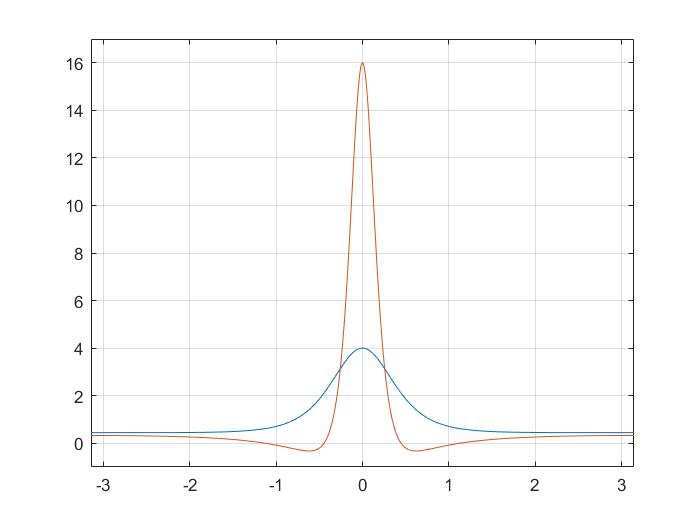}}

$$Figure \; 2$$

As $r$ gets closer to 1 the graph becomes narrower and the peak higher and, not surprisingly, tends to Dirac $\delta$-function. In that sense the $Q$-kernel behaves almost identically with the Poisson kernel except of the fact that it is not non-negative. Before going to differences here we would like to demonstrate more similarities between the two integral representation formulas (4) and (6). Let us consider one of the simplest possible harmonic functions $u=x=r\cos\theta$. The boundary value of this function on the unit circle is obviously $f(\theta)=\cos\theta$ (here and later we write $f(\theta)$ instead of $f(e^{i\theta})$, i.e. we identify the unit circle with the interval $[0,2\pi]$). Because the function $u$ is harmonic we would obviously have 
$$r\cos\theta=\frac{1}{2\pi}\int_{-\pi}^{ \pi}\cos\phi \frac{1-r^2}{1-2r\cos(\theta-\phi)+r^2} d\phi$$
and (because $u(0)=0),$
$$r\cos\theta=\frac{2}{\pi}\int_0^1\int_0^{2\pi}\rho\cos\phi\frac{1-2r\rho\cos(\theta-\phi)+r^2\rho^2\cos2(\theta-\phi)}{(1-2r\rho\cos(\theta-\phi)+r^2\rho^2)^2}d\phi \rho d\rho.$$

In other words, formulas (4) and (6) produce (reconstruct) the exact same function. This would be the case any time we use these formulas for functions harmonic in the unit disk. We will demonstrate this fact also graphically below in the case of another simple harmonic function $u(x,y)=x^2-y^2=r^2\cos2\theta$ with the boundary value $f(\theta)=\cos2\theta$. 

\centerline{\includegraphics[width=6in]{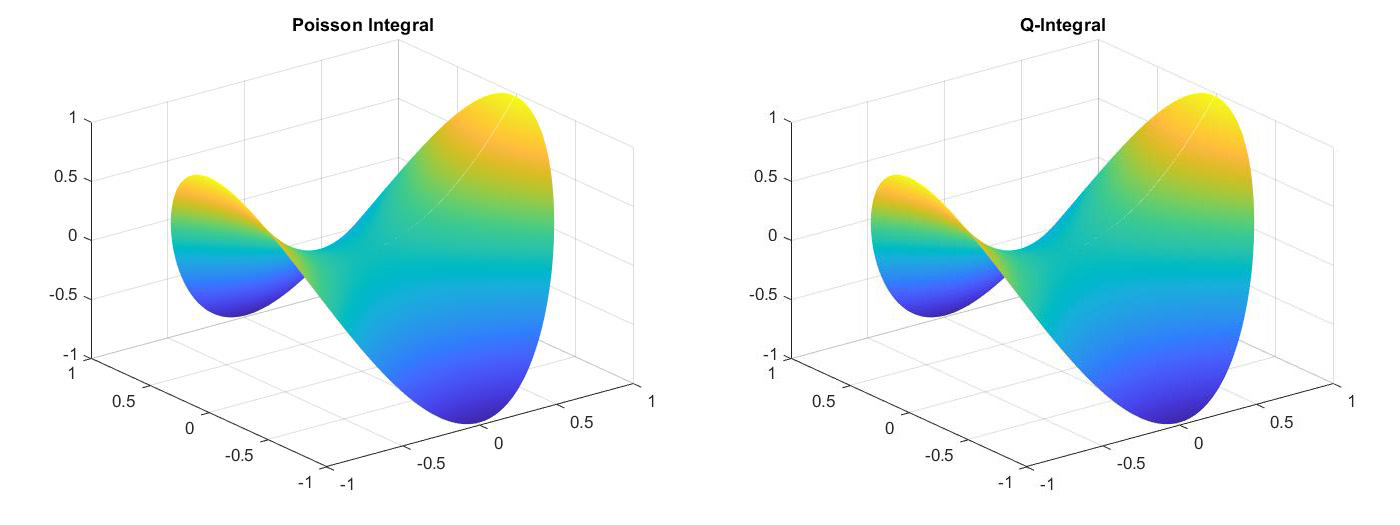}} 

$$Figure \; 3$$

As it is clearly seen the graphs match perfectly not just in shape and form but also numerically.\footnote{Numeric outputs for these and all the following graphs are available upon request by writing to the first author}

Next we will start concentrating on essential differences between these two integral representations and in order to do that we will need some more general results about the kernel $Q$ and the integral formula (6). For that purpose it is convenient to consider an integral operator

\begin{equation}
Tf(r,\theta)=\int_D f(\rho,\phi)Q(r\rho,\theta-\phi)\rho d\rho d\phi 
\end{equation}
where the function $f$ belongs to the Lebesgue space $L^2=L^2(D, dxdy).$ Then the following result is a very special case of the Theorem 7.3 from [2]:

\begin{theorem}
The operator $T$ is a bounded projection from $L^2$ to $hA^2$ and the norm of this operator is $\leq1$:  $||Tf||_{hA^2}\leq||f||_{L^2}$
\end{theorem}

This means, in particular, that for any square integrable function $f$ (harmonic or not) the result of its convolution with the kernel $Q$ is a harmonic function which belongs to the class $hA^2$. Calculating integral (8) for non-harmonic functions explicitly is very difficult except when $f$ is a polynomial. For a software allowing to do it symbolically using Mathematca see [1]. For any other types of functions, especially non-continuous functions, it is basically impossible. In this article we, for the first time,  have collected a number of cases where we have calculated this integral numerically and graphed them.

Our first example will be the $Q$-integral of the characteristic function of the disk with radius $r=1/4$ centered at the origin.
\begin{equation}
    \int_0^{1/4}\int_0^{2\pi}Q(r\rho,\theta-\phi)\rho d\phi d\rho.
\end{equation}
The output function is depicted below:

\centerline{\includegraphics[width=4in]{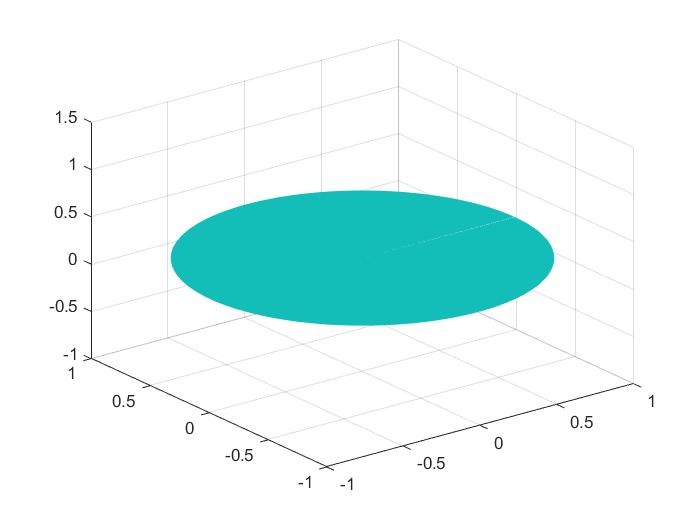}}

$$Figure \;4$$

The height of the disk is about $\pi/16\approx 0.193.$ This result seems a little surprising but makes sense because the resulting  function should be harmonic and also makes sense from a physical point of view. We will return to this issue later in the next section.

In the next two examples we first calculate the $Q$-integral of the characteristic function of the polar rectangle $[1/4,1/2]\times[0,\pi/4]$ 
$$\int_{1/4}^{1/2}\int_0^{\pi/4}Q(r\rho,\theta-\phi)\rho d\phi d\rho.$$
and the next graph is the $Q$-integral of the above characteristic function plus the characteristic function of the polar rectangle $[0.6,0.8]\times[5\pi/6, \pi]$:
$$\int_{1/4}^{1/2}\int_0^{\pi/4}Q(r\rho,\theta-\phi)\rho d\phi d\rho+ \int_{0.6}^{0.8}\int_{5\pi/6}^{\pi}Q(r\rho,\theta-\phi)\rho d\phi d\rho.$$
The graphs are presented below side by side for comparison.

\centerline{\includegraphics[width=6in]{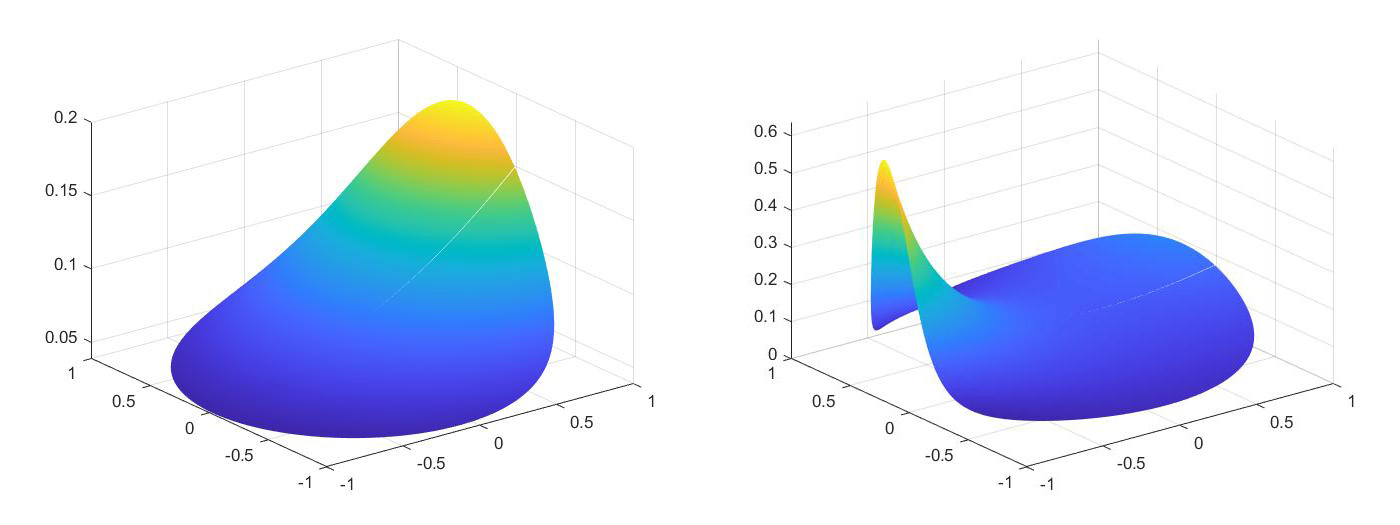}}

$$Figure \; 5$$

By Theorem 2 both functions are clearly harmonic. It is also obvious from comparison that the second function has more "mass" because the input is bigger. If the highest point of the first function is just about 0.17 then for the second one it is about 0.5. 

Our next example is a function that goes to infinity near the boundary of the unit disk. The function
$$f(r,\theta)=\frac{\cos\theta}{(1-r)^{1/4}}$$
is not harmonic but clearly belongs to $L^2(D)$. We take the $Q$-transform of that function multiplied by the characteristic function of the "rectangle" $[3/4,1]\times[-\pi/6,\pi/6]$:
$$\int_{3/4}^1\int_{-\pi/6}^{\pi/6}\frac{\cos\phi}{(1-\rho)^{1/4}}Q(r\rho,\theta-\phi)\rho d\phi d\rho.$$

\centerline{\includegraphics[width=4in]{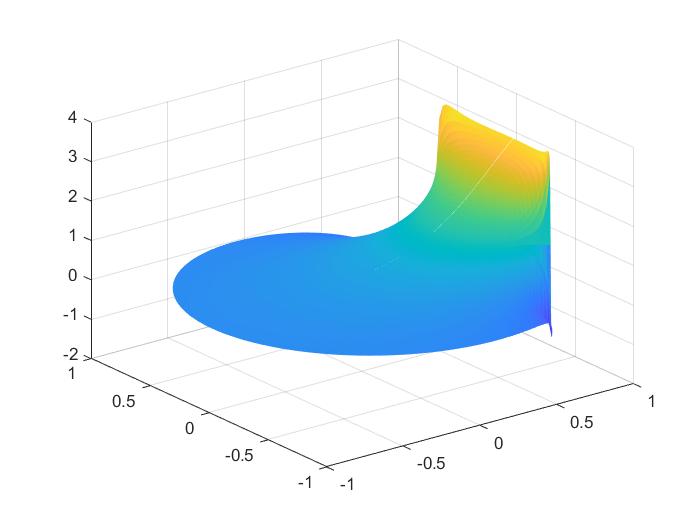}}

$$Figure \; 6$$

The resulting function grows near the point $\theta=0$ and the picture is necessarily cut off because of that. 

And in our final example we would like to show the behaviour of the $Q$-transform of another $L^2$ function going to infinity near the boundary which is the combination of the previous integral and a similar function that tends to $-\infty$ on the opposite side of the disk:

$$\int_{3/4}^1\int_{-\pi/6}^{\pi/6}\frac{\cos\phi}{(1-\rho)^{1/4}}Q(r\rho,\theta-\phi)\rho d\phi d\rho+\int_{7/8}^1\int_{5\pi/6}^{\pi}\frac{\cos\phi}{(1-\rho)^{3/8}}Q(r\rho,\theta-\phi)\rho d\phi d\rho.$$

You can see from the figure bellow that the behaviour of this graph almost completely matches the behaviour of the previous one near $\theta=0$. On the other hand on around the interval $[5\pi/6,\pi]$ it has the complete opposite behaviour. As in the previous graph we had to cut off the graph to allow the program work.

\centerline{\includegraphics[width=4in]{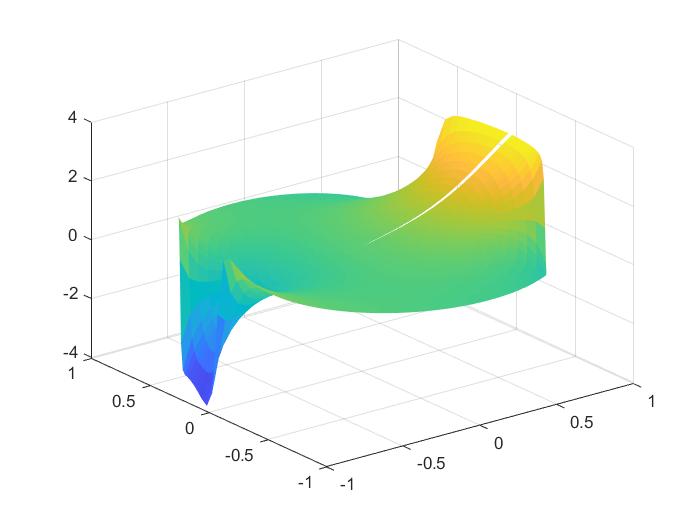}}

$$Figure \; 7$$

\section{The steady-state heat equation}

Let us return to the classical Dirichlet problem for the Laplace's equation in the disk discussed at the beginning of the previous section:
 \begin{equation}
 \Delta u=0, \quad u|_{T}=f,
 \end{equation}
where $f\in C(T)$. As we have mentioned earlier the solution to this boundary problem is given by the Poisson integral
\begin{equation}
    u(r,\theta)=\frac{1}{2\pi}\int_{-\pi}^{\pi}f(\phi)\frac{1-r^2}{1-2r\cos(\theta-\phi)+r^2}d\phi.
\end{equation}
However, the formula (11) is valid for much larger class of functions $f(\theta)$ but we restricted our interest to the case $f\in L^2(T)$ to avoid discussion of fine properties of harmonic functions at the boundary. Among other things the boundary problem (10) solves the following "steady-state" (or time independent) heat transfer problem: Assume that we establish and maintain certain temperature on the unit circle which is described by the function $f(\theta)$. How the temperature will be distributed at each point of the disk. Formula (11) gives exact answer to that question. It is also important to remind that the resulting function $u(r,\theta)$ is harmonic and, in particular, satisfies the minimum and maximum principles for harmonic functions: Function $u$ can reach its maximum or minimum values only on the boundary or that function is constant.  

Even though the results described above (and even their very far reaching generalizations) are known for a very long time we find it useful to provide some illustrations of behaviour of solutions of the boundary problem (10). 

In our first example the function $f$ is just the characteristic function of the integral $[-\pi/6,\pi/6]$ (otherwise called the harmonic measure of that interval):

$$  u(r,\theta)=\frac{1}{2\pi}\int_{-\pi/6}^{\pi/6}\frac{1-r^2}{1-2r\cos(\theta-\phi)+r^2}d\phi$$

\centerline{\includegraphics[width=4in]{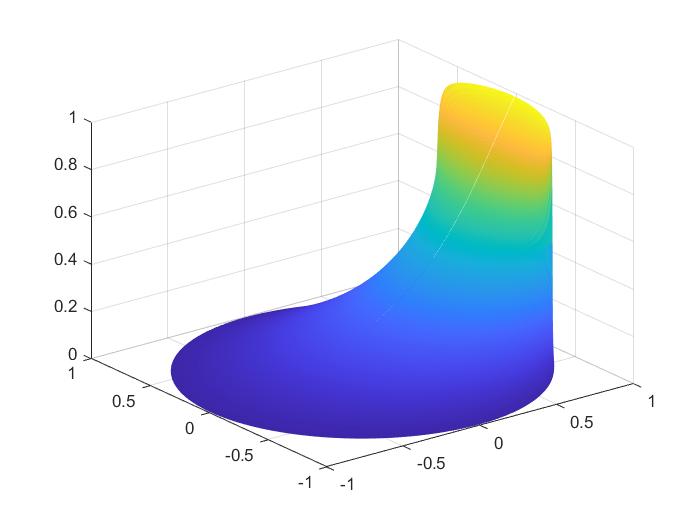}}

$$Figure \; 8$$

As we clearly see the resulting function reaches its maximum value 1 on some interval around $\theta=0$ and is zero on the compliment of that interval. Disregarding some complications at the end points $\pm \pi/6$ inside the disk the resulting function is clearly harmonic.

Now let us consider a different physical problem. Suppose we are heating not just boundary (or part) of the disk but a whole layer close to the boundary. For example, that might be a polar rectangle of the width 0.1 $R=[0.9,1]\times[-\pi/6,\pi/6]$. It is obvious that the Poisson formula (11) will not work in this case. It is also clear from physical considerations that the difference should not be crucial. So, let us see if the $Q$-transform can do the job here. In other words, let us calculate and graph the function 
\begin{equation}
\frac{2}{\pi}\int_{0.9}^1\int_{-\pi/6}^{\pi/6}\frac{1-2r\rho\cos(\theta-\phi)+r^2\rho^2\cos2(\theta-\phi)}{(1-2r\rho\cos(\theta-\phi)+r^2\rho^2)^2}d\phi \rho d\rho.
\end {equation}

The resulting function is depicted in the graph below from two different angles:

\centerline{\includegraphics[width=5in]{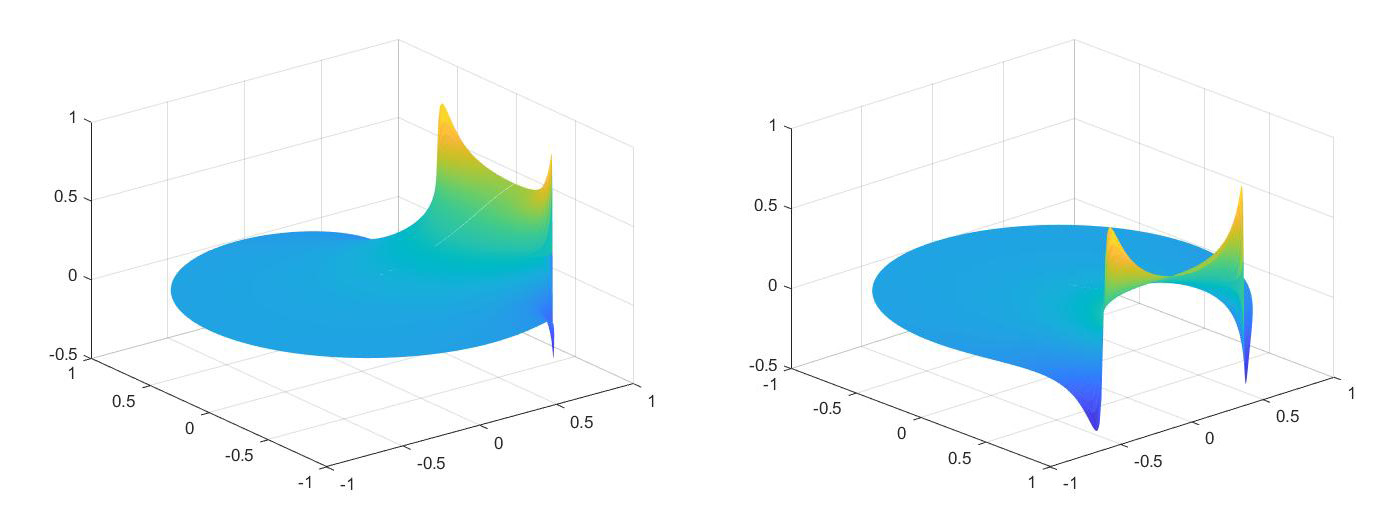}}

$$Figure \; 9$$

Let us now compare the form and numeric values of functions in Figures 8 and 9.  First of all we notice that the shape of the graphs are almost identical away from "critical" rectangle $R$. However, the numeric values of the $Q$-integral are just about half of the Poisson integral. When we approach to the rectangle $R$ differences are becoming even more significant and near the very edge a big drop in values happens: at some small parts in $R$ the values even become negative. That is clearly visible in the  right-hand side picture in Figure 9.  

So, the first conclusion seems to be that this approach to the solution of the "non-standard" problem is unlikely to be given by the integral (12). However, let us continue with more examples and compare the resulting graphs.In our second example the function $f$ in (11) is $|\theta|$ and we present it in two different views:

\centerline{\includegraphics[width=6in]{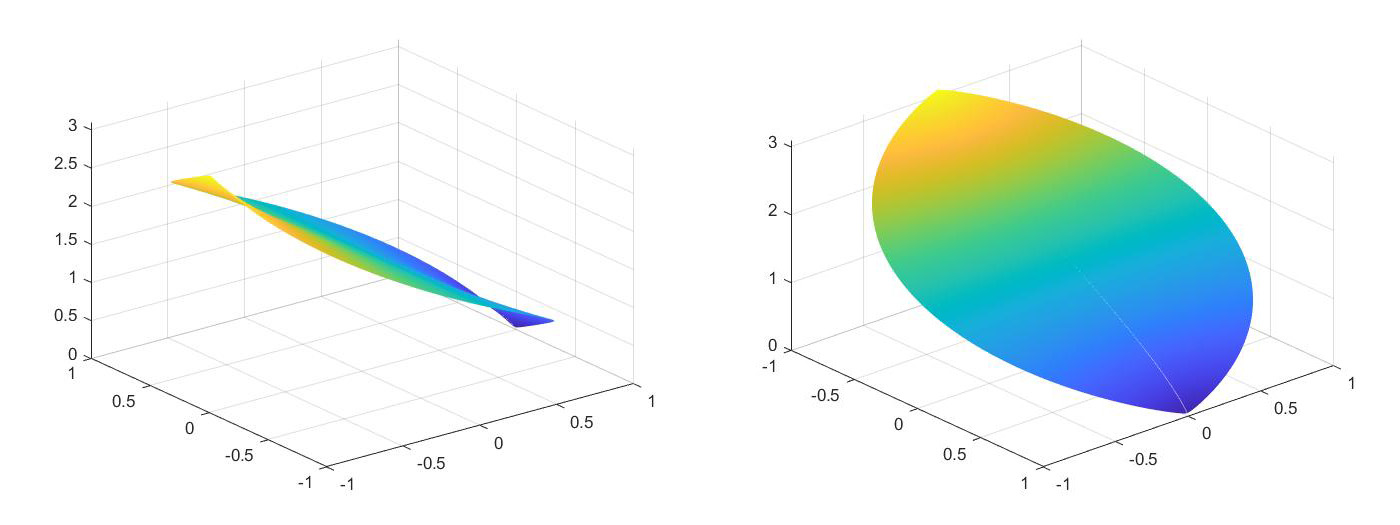}}

$$Figure \; 10$$

Next we will try to pose the corresponding question similar to the previous case. Suppose a whole layer of the disk $D$ is heated with the width 0.1, say. What will be the heat distribution throughout the whole disk. Let us try again to use the $Q$-transform:   

$$\frac{2}{\pi}\int_{0.9}^1\int_{-\pi}^{\pi}\frac{1-2r\rho\cos(\theta-\phi)+r^2\rho^2\cos2(\theta-\phi)}{(1-2r\rho\cos(\theta-\phi)+r^2\rho^2)^2}|\phi|\rho^2d\phi d\rho.$$

The resulting graphs are below:

\centerline{\includegraphics[width=6in]{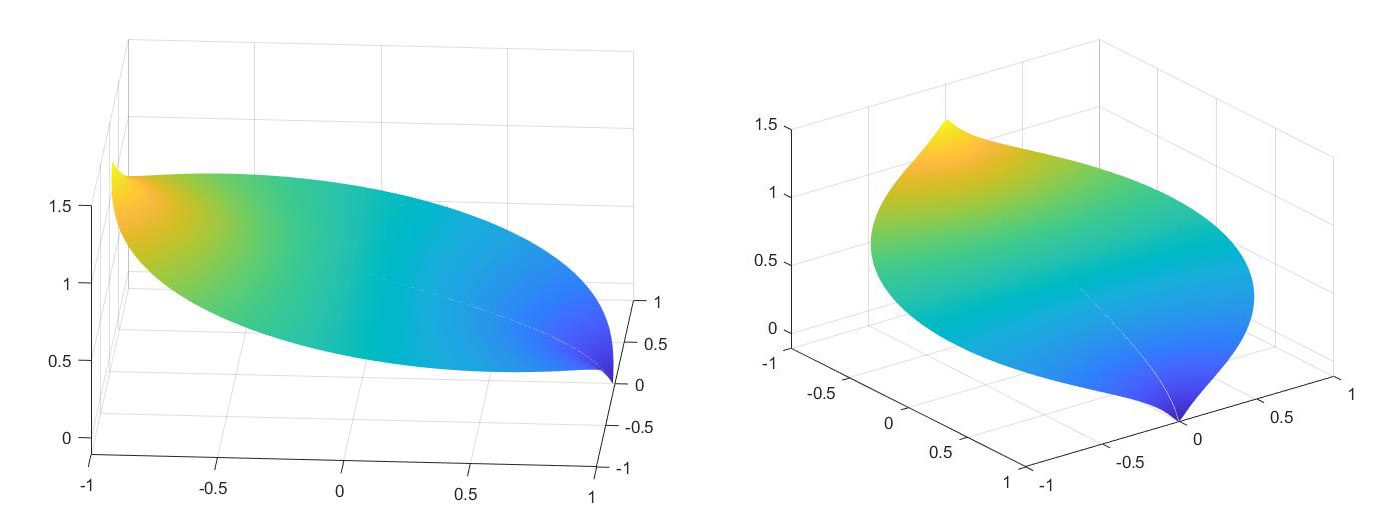}}

$$Figure \; 11$$

Comparison of Figures 10 and 11 gives us a much better match. The graphs are clearly very similar (although certainly not identical). The numeric values in the second case are again roughly the half of the first one. 

Generally speaking, when comparing Poisson integrals of certain functions on the unit circle and $Q$-transforms of corresponding functions on a narrow strip about the circle we see that sometimes there are substantial differences and in many other cases general similarities in shapes of corresponding graphs. In both situations, of course, numerically they are different but seems that follow the same pattern: Poisson integrals (at least for positive functions) are greater than the $Q$-transforms. To confirm this assertion we will present some more examples.  In the next graph we put next to each other Poisson integral of the function $f(\theta)=\theta^2$ on interval $[-\pi/6,\pi/6]$ and the $Q$-integral of the corresponding function $f(r,\theta)=r\theta^2$ on rectangle $[0.9,1]\times[-\pi/6,\pi/6]$:
$$\frac{1}{2\pi}\int_{-\pi/6}^{\pi/6}P_r(\theta-\phi)\phi^2 d\phi, \quad \frac{2}{\pi}\int_{0.9}^1\int_{-\pi/6}^{\pi/6} Q(r\rho, \theta-\phi)\phi^2\rho^2 d\phi d\rho$$
The resulting graphs are below:

\centerline{\includegraphics[width=6in]{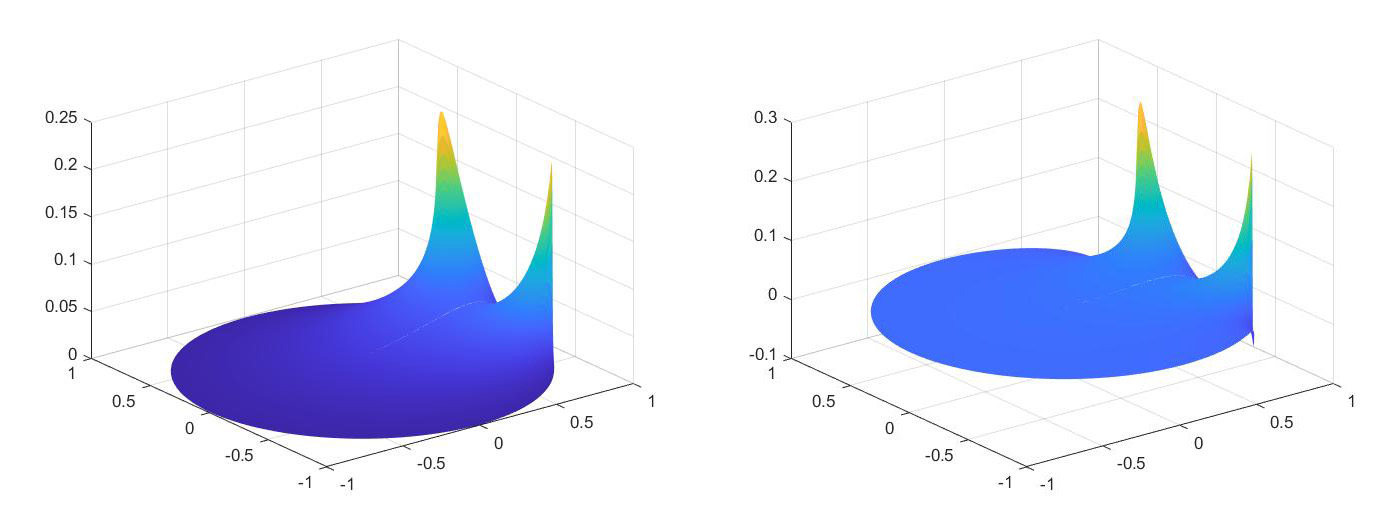}}

$$Figure \; 12$$

Here we clearly see the similarities of the behaviour of both harmonic functions and the differences in numeric values. We can see a very similar pattern also in the next example where we have taken the Poisson integral of the function $\sin\theta$ over interval $[0,\pi$ and the $Q$-transform of the corresponding function $r\sin\theta$ over the rectangle $[0.9,1]\times[0,\pi]$:

$$\frac{1}{2\pi}\int_0^{\pi}P_r(\theta-\phi)\sin\phi d\phi, \quad \frac{2}{\pi}\int_{0.9}^1\int_0^{\pi} Q(r\rho, \theta-\phi)\sin\phi\rho^2 d\phi d\rho$$

\centerline{\includegraphics[width=6in]{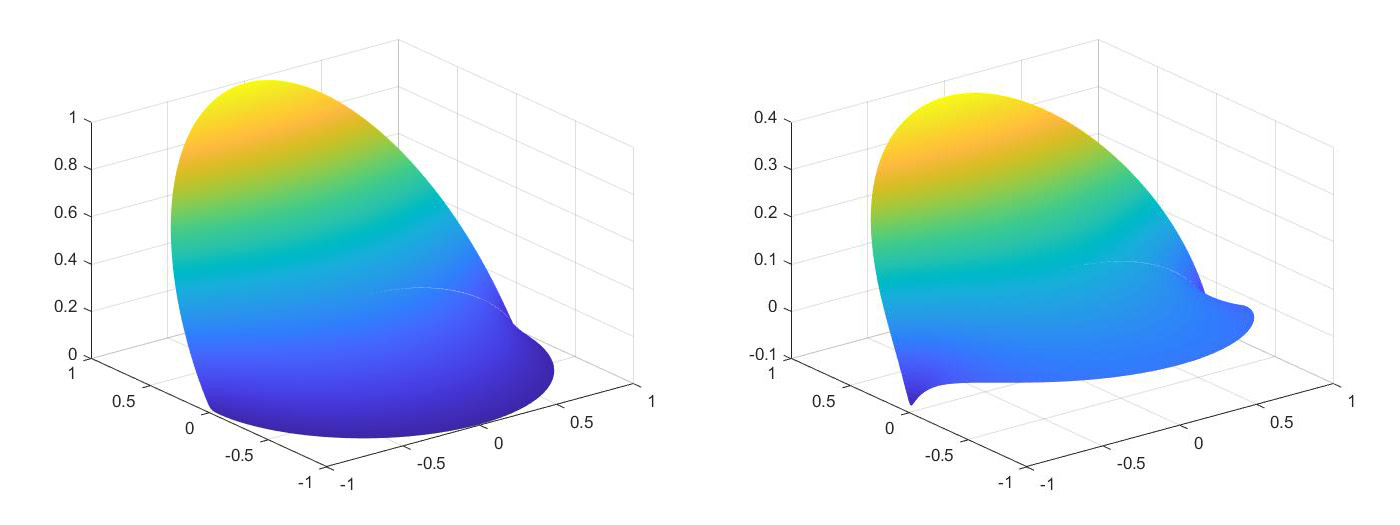}}

$$Figure \; 13$$

And again we see pretty similar pattern except the fact that the second function takes some negative values too near the critical points $\theta=0, \pi.$

Our last example in this part is a function with integrable singularity, namely the logarithmic function:

$$\frac{1}{2\pi}\int_0^{\pi}P_r(\theta-\phi)|\ln|\phi|| d\phi, \quad \frac{2}{\pi}\int_{0.9}^1\int_0^{\pi} Q(r\rho, \theta-\phi)|\ln|\phi||\rho^2 d\phi d\rho$$
and the graphs  below again show a similar pattern as in the previous examples:

\centerline{\includegraphics[width=6in]{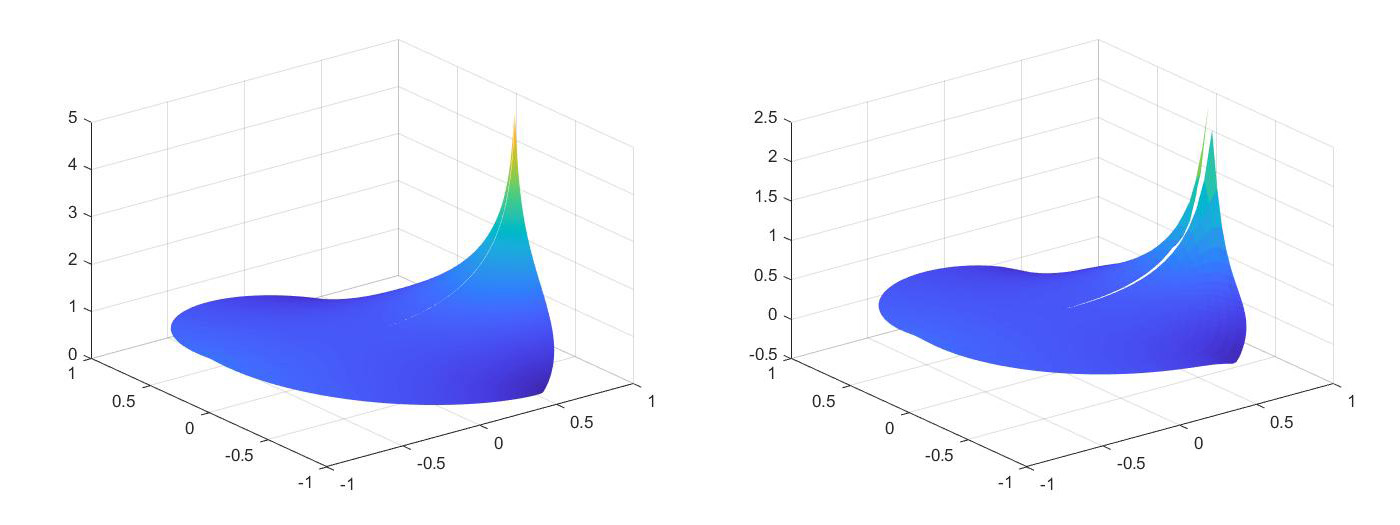}}

$$Figure \; 14$$

At this point we would like to return to some examples from Section 2. In these examples comparison with Poisson integrals is no longer possible because the functions we have used there were supported inside the disk either completely or had significant "mass" inside the unit disk.

Let us look back at the integral (9) and the corresponding graph in Figure 4 first. Then we can see that the result conforms (at least in general terms) with physical intuition. Really, in that integral if we view the characteristic function of the disk $r\leq 1/4$ as a source of heat of "intensity" one, then the common sense tells us that eventually  the heat will be spread uniformly throughout the whole disk $r\leq1$ but the temperature will be lower than initial temperature 1. Pay attention that in this case it would be impossible to apply Poisson integral to come to the same conclusion. 

We will come to similar conclusions if we look at integrals that resulted in graphs depicted in Figures 5, 6, and 7. It could be claimed with reasonable certainty  that these graphs represent the heat distribution in the disk with the corresponding  "heat sources".

We would like to add one more example along these lines. The function which we view as a heat source is $f(r,\theta)=10e^{-10(r-0.5)^2}\cos\theta$ restricted to the "rectangle" $[0.3,0.7]\times[-\pi/6,\pi/6]$. This function is supported inside the disk and, again, cannot be handled with Poisson integral. The resulting graph represents a harmonic function that has growth tendency in the direction of the boundary where the presumed heat source is concentrated. We present below that graph along with the function $f(r,\theta)$ to the right.

\centerline{\includegraphics[width=6in]{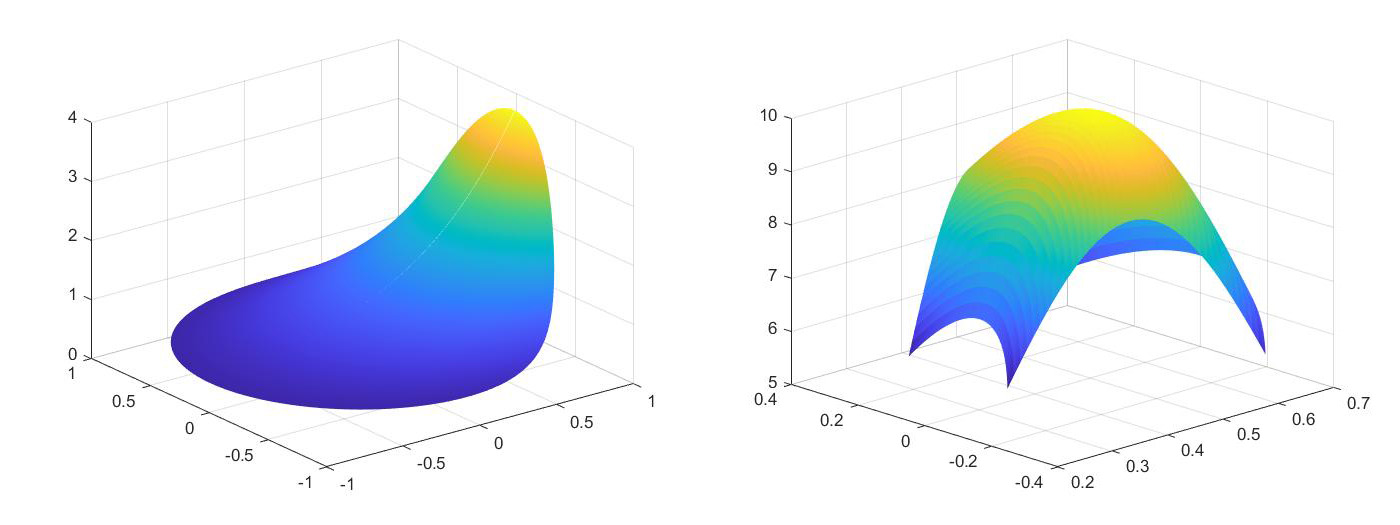}}

$$Figure \; 15$$

Based on all the previous examples we would like to formulate a plausible conjecture:\\

\noindent\textbf {Conjecture} {\it Suppose a metallic disk is getting  heated by a source described by the function $f(r,\theta)$. Then the steady-state heat distribution through the disk, assuming there are no outside factors, is given by the integral (8) modulo a multiplicative constant depending on physical properties of the disk.} \\

 Mathematical proof seems to be elusive at this juncture but it could be confirmed or rejected also experimentally.\\

\noindent CONCLUDING REMARK. The coding, numeric calculations, and graphing part of this article was done completely by the second author who is a student at Glendale Community College.\\

\noindent REFERENCES\\[10pt]

\noindent 1. S.Axler , P. Bourdon, W.Ramey,  "Harmonic Function Theory", Springer Verlag, New York, Berlin, 1992\\
2. A.Djrbashian, F.Shamoian, "Topics in the Theory $A_\alpha^p$ Spaces", Teubner Verlag, Leipzig, 1988\\
3. M.M.Djrbashian, "On the problem of representing analytic functions", Soobshch. Inst.Math. Mech.Acad.Nauk Arm.SSR, N2(1948), 3-40.\\ 
4. P.L.Duren, "Theory of $H^p$ Spaces", Academic Press 1970\\
5. P.L.Duren, A.Schuster, "Bergman Spaces", AMS, Providence RI, 2004\\
6. H.Hadenmalm, B.Korenblum, K.Zhu, "Theory of Bergman Spaces", Springer Verlag, New York, Berlin, 2000\\

Ashot Djrbashian

Glendale Community College, Glendale CA

E-mail: ashotd@glendale.edu

Armen Arakelyan

Glendale Community College, Glendale, CA

E-mail: aarakel632@student.glendale.edu

\end{document}